\documentclass[journal,10pt]{IEEEtran}  
\IEEEoverridecommandlockouts
\usepackage{cite}
\usepackage{indentfirst}
\usepackage{graphicx}
\usepackage{changepage}
\usepackage{stfloats}
\usepackage{amsfonts,amssymb}
\usepackage{bm}
\usepackage{diagbox}
\usepackage{algorithm}
\usepackage{algorithmic}
\usepackage{amsmath}
\usepackage{amsmath,amssymb,amsfonts}
\usepackage{times}
\usepackage{url}
\usepackage{cite}
\usepackage{textcomp}
\usepackage{xcolor}
\usepackage{color}
\usepackage{setspace}
\usepackage{enumitem}
\usepackage{float}
\usepackage{diagbox}
\usepackage[T1]{fontenc}
\usepackage[utf8]{inputenc}
\usepackage{authblk}
\usepackage{threeparttable}
\usepackage{booktabs}
\usepackage{footnote}
\usepackage{graphicx}
\usepackage{pifont}
\usepackage{subfigure}
\usepackage{mathrsfs}
\usepackage{makecell}
\usepackage{array}
\usepackage{booktabs}
\usepackage{lipsum}
\usepackage{multirow}
\usepackage{adjustbox}
\usepackage{bbding}
\usepackage{tabularx}
\usepackage{colortbl}

\begin{document}
\title{Cooperative ISAC Networks: Opportunities and Challenges}

\author{
	Kaitao Meng, \textit{Member, IEEE}, Christos Masouros, \textit{Fellow, IEEE}  Athina P. Petropulu, \textit{Fellow, IEEE}, and Lajos Hanzo, \textit{Life Fellow, IEEE} 
	\thanks{{K. Meng, and C. Masouros are with the Department of Electronic and Electrical Engineering, University College London, London, UK. }{(emails: \{kaitao.meng, c.masouros\}@ucl.ac.uk).} A. P. Petropulu is with the Department of Electrical and Computer Engineering, Rutgers University, Piscataway, NJ 08901 USA (email: athinap@rustlers.edu). L. Hanzo is with School of Electronics and Computer Science, University of Southampton, SO17 1BJ Southampton, UK (email: lh@ecs.soton.ac.uk)}.
}

\maketitle


\begin{abstract}
	The integration of sensing and communication (ISAC) emerges as a cornerstone technology for the sixth generation era, seamlessly incorporating sensing functionality into wireless networks as a native capability. The main challenges in efficient ISAC are constituted by its limited sensing and communication (S\&C) coverage, as well as severe inter-cell interference. Network-level ISAC relying on multi-cell cooperation is capable of effectively expanding both the S\&C coverage and of providing extra degrees of freedom (DoF) for realizing increased integration gains between S\&C. In this work, we provide new considerations for ISAC networks, including new metrics, the optimization of the DoF, cooperation regimes, and highlight new S\&C tradeoffs. Then, we discuss a suite of cooperative S\&C architectures both at the task, as well as data, and signal levels. Furthermore, the interplay between S\&C at the network level is investigated and promising research directions are outlined.
\end{abstract}


\section{Introduction}

\par
ISAC (integrated sensing and communication) is a technique that utilizes the same frequency, waveform, and infrastructure to simultaneously achieve data transmission and extract target information from scattered echoes, thereby substantially improving spectrum, energy, and cost efficiency in sensing and communication (S\&C) functionalities \cite{Lu2024Integrated}. ISAC has recently attracted vibrant research interests as a promising next-generation networking paradigm \cite{Zhang2021OverviewSignal}. In the literature, most of the existing studies on this topic focus primarily on the link/system level ISAC design \cite{Liu2022Integrated}. However, these studies overlook some critical challenges in practice, such as the severe inter-cell interference that inherently limits performance, the restricted coverage attributable to potential obstruction and attenuation experienced at high frequencies, as well as the disparity of the achievable S\&C range due to the two-hop pathloss in the sensing process.

\begin{figure}[t]
	\centering
	\includegraphics[width=8cm]{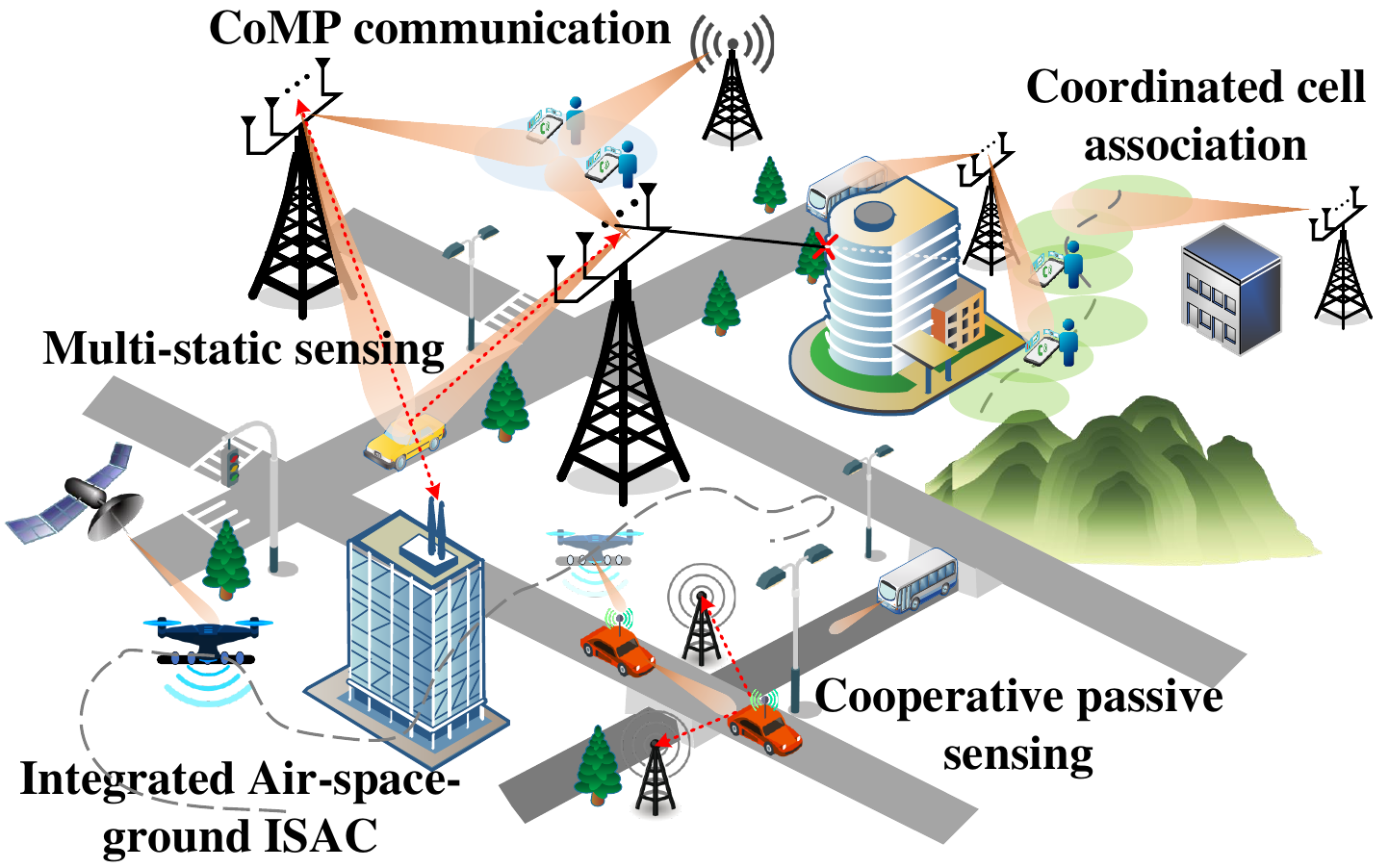}
	\vspace{-3mm}
	\caption{Network-level ISAC scenarios.}
	\label{figure1}
\end{figure}

Network-level ISAC refers to the collaboration of multiple ISAC transceivers across different cells to enhance both wireless communication and sensing. This approach aims to deliver high-throughput, ultra-reliable, low-latency communication alongside ultra-precise, high-resolution, and robust sensing capabilities \cite{Li2023TowardsSeamless, Meng2024Cooperative}. As shown in Fig.~\ref{figure1}, on the sensing side, the ISAC network can cover larger surveillance areas than single-cell ISAC, while providing wider sensing angles and capturing richer sensing information \cite{Li2023TowardsSeamless}. This is because, in addition to radar echoes generated within a single cell, each ISAC base station (BS) may also receive the target-reflected signals transmitted by other BSs or users, thereby forming multi-static sensing \cite{Shi2022Device}. On the communication side, multiple ISAC transceivers can collaborate to establish connections with several users through advanced coordinated multi-point (CoMP) transmission techniques for managing inter-cell interference \cite{Hosseini2016Stochastic}. Therefore, through strategic adaptation of S\&C cooperation at the task, data, and signal levels, ISAC networks offer new opportunities for refining the allocation of network-level resources, leading to enhanced signal power, controllable interference management, improved coverage quality, and enhanced mutual assistance between S\&C capabilities.

The advantages of cooperative ISAC networks come at increased signaling overhead and resource consumption due to the need for information exchange between transceivers and control unit centres \cite{Liu2022Integrated}. Thus, it is critical to design a cooperation framework based on specific task requirements for striking a tradeoff between the S\&C performance gains and control signalling costs. In addition, networked ISAC also faces new technical challenges in wireless resource allocation and user/target scheduling. These challenges arise from performance characterization and conflicting requirements in S\&C collaboration at the network level. It is essential to accurately characterize the average S\&C performance for optimizing the ISAC networks. This involves addressing the challenges posed by uncertainties such as channel fading and the mobility of users and targets. 
Upon evolving from communication-only to ISAC networks, several questions have to be addressed. For instance, do traditional network-level coordination techniques tailored for communication-only networks apply to the ISAC paradigm? How can performance metrics, coordination, and resource allocation be tailored for improving the joint S\&C performance at the network level? 
Based on the above questions, we offer a comprehensive overview of network-level ISAC, pointing out key challenges, exploring potential solutions, and identifying open research directions. 

\section{New Insights For Network-Level ISAC}
\label{KeyTechnology}
\subsection{New Considerations}
\label{Newconsideration}
In this section, we provide several new considerations when analyzing, evaluating, and optimizing ISAC networks, including metrics, optimization DoF, constraints, the framework, and S\&C tradeoffs, as outlined in Table \ref{Table1}.

\begin{table*}[t] 
	\centering
	\small
	\caption{Comparison between system-level ISAC and network-level ISAC.} 
	\label{Table1}
	\begin{center}
		\begin{tabular}{ |p l | l | l | l| l | l|}
			\hline 
			\multicolumn{1}{|c|}{{  ISAC Levels }}
			 & {\makecell[c] {{Performance metrics}}}   & {\makecell[c] {{Optimization DoF}}}   & {\makecell[c] {{Resource}\\{constraints}}} & {\makecell[c] {{Cooperation}\\{frameworks}}}  & {\makecell[c] { {{Tradeoffs}}}}  \\ 
			\hline
			\multicolumn{1}{|c|}{{  \makecell[c]{{System-level}\\{ISAC}}} } & {\makecell[c]{SINR, achievable rate, \\outage probability,\\CRLB, sensing rate,\\MMSE}}  & {\makecell[c] {Resource allocation,\\{antenna selection,}\\{beamforming}}}   & {\makecell[c]{Power, time,\\ energy constraints}} & {\makecell[c]{synergy between\\S\&C functionalities,\\user/target/BS\\ collaboration}} & {\makecell[c]{{Deterministic and}\\{random signals,}\\{channel correlation}}}  \\ 
			\hline
			\multicolumn{1}{|c|}{{  \makecell[c]{{Network-level}\\{ISAC}}} } & {\makecell[c]{{CRLB coverage}\\probability, joint \\S\&C ASE, networked\\energy efficiency}}  & {\makecell[c]{BS density,\\cooperative cluster\\size, cell association}}   & {\makecell[c]{{Joint S\&C backhaul}\\{capacity, BS load,}\\{interference nulling}\\{DoF}}} & {\makecell[c]{{BS topology}\\{design, distributed}\\{dynamic cluster}}} & {\makecell[c]{{Correlation between}\\{signals transmitted by}\\{different BSs},\\{deployment geometry}}}  \\ 
			\hline
		\end{tabular}
	\end{center}
\end{table*} 

\subsubsection{Network-level Performance Metrics}
To optimize and balance the S\&C performance across the entire ISAC network, new metrics are needed, which are different from the popular link-level ISAC metrics. For instance, in contrast to the performance analysis of communication-only networks based on the distribution of signal-to-interference-plus-noise ratio (SINR), a more applicable approach for evaluating sensing coverage may lie in defining metrics aligned with estimation theory, such as the likelihood of the average Cramér-Rao lower bound (CRLB) of the network not exceeding a predefined threshold. Here, CRLB constitutes a measure of the lowest possible variance of an estimator, helping to evaluate the accuracy of sensing operations. However, this kind of metric has not yet been investigated in the literature. Moreover, how to describe the spectral efficiency of ISAC networks is also a new and challenging problem to be tackled due to the difficulty of quantifying the average information of sensing operations of the ISAC network in terms of bits. Most recently, the authors in \cite{Meng2024Cooperative} proposed a unified area spectral efficiency (ASE) to describe the average S\&C performance of the ISAC networks, where the unified ASE is a metric defined to evaluate the spectrum efficiency of both sensing and communication in terms of the data rate in bits/sec/Hz/km$^2$.

\subsubsection{Network-level Optimization of DoF}
In addition to optimizing the resources at a single ISAC BS, to effectively balance the S\&C performance at the network level, it is necessary to optimize additional resource variables, including the BS density, per-BS frequency/power allocation, the cluster sizes of cooperative S\&C BS sets, and cell association. By doing so, cooperative transmission and sensing can strike a more flexible tradeoff between the S\&C performance at the network level \cite{Meng2024Cooperative}.

\subsubsection{Network-level Resource Constraints}
\label{ResourceConstraints}
Considering the high demand for information exchange between ISAC BSs when achieving cooperative S\&C, new constraints must be introduced at the network level. In a cooperative S\&C network, the cluster sizes of cooperative sensing/communication are limited by the backhaul link capacity, as data volume for information sharing and echo signal collection increases with cluster size \cite{Ghimire2015Revisiting}. Additionally, due to the limited number of transmit antennas, the constraints on the maximum user/target load and the DoF in spatial resource allocation for S\&C interference nulling and multiplexing gain improvement become crucial in the family of ISAC networks \cite{meng2023network}.

\subsubsection{Network-level Cooperation Framework}
In ISAC networks, diverse information having time-variant requirements has to be exchanged and sent back to the control centre, e.g, raw sensory data, target channel state information (CSI), updated gradient information of neural networks, semantic information, etc.
In this case, the BS connection and topology significantly affect resource scheduling and collaboration efficiency, especially in dynamic ISAC networks. This also facilitates the design of advanced network resource management techniques based on real-time traffic and sensing requirements, such as dynamic spectrum allocation using cognitive radio technologies. The ISAC network architecture and MAC (medium access control) layer protocols have to accommodate not only the high volume of data transmission, but also meet the specific demands of distributed sensing, such as strict time synchronization and sensory data timeliness requirements \cite{Zhang2021OverviewSignal}. In general, it is challenging for conventional communication-only networks, e.g.,  cloud radio access network (C-RAN) and cell-free networks, to guarantee the sensing performance of ISAC networks. Hence, substantial changes are necessary for flawlessly integrating sensing into the cellular network architecture.

\subsubsection{Network-level Tradeoffs}
Emerging network-level factors, such as BS deployment and the correlation among signals transmitted by different BSs, play a crucial role in striking a balance between S\&C performance. Specifically, ensuring orthogonality among signals     
arriving from different BSs is critical for distributed radar sensing performance \cite{Meng2024Cooperative}. However, this constraint may impose the unintended consequence of reducing the transmission rate of each BS, as it contradicts the deliberate randomness of signals for achieving Shannon capacity. In addition, BS selection/placement entails tradeoffs that influence S\&C coverage and service quality. For instance, communication-optimal deployment typically prioritizes distance-based assignment between users and transceivers, while sensing-optimal deployment must also consider geometric factors such as diversity in directions between targets and transceivers \cite{Yang2023Deployment}. The above-mentioned tradeoffs mainly arise from the involvement of multiple BSs in S\&C tasks, which is different from that observed at the link-level ISAC \cite{Liu2023Seventy}, such as deterministic and random signal transmission as well as the correlation between user channel and target channel.

\subsection{New Opportunities}

The use of network-level ISAC offers new opportunities for improved S\&C performance, overcoming the limitations of conventional link/system level ISAC. In this section, we outline some typical scenarios where ISAC networks can provide significant benefits.

\subsubsection{Space-Air-Ground ISAC Networks}
Space-air-ground integrated (SAGI) ISAC networks rely on satellite systems, aerial platforms, and terrestrial infrastructures. They constitute a promising architecture having seamless S\&C coverage \cite{Yang2023Deployment}, as shown in Fig.~\ref{figure1}. SAGI ISAC networks can gather comprehensive status information about users, targets, and the surrounding environment, while transmitting data between transceivers and users employing unified signals. They also assist in enhancing propagation, creating line-of-sight links and improving network coverage. This offers profound opportunities for dynamically allocating resources and making comprehensive decisions according to the sensing results. 

\subsubsection{Multi-modal Sensing Information Transmission and Fusion}
At the network level, aggregating and fusing the multi-modal sensing results gleaned from ISAC BSs and other nodes equipped with diverse sensor equipment can significantly enhance the robustness of network sensing \cite{li2022multi}, thus further bolstering communication performance. The integration of multi-modal sensing information, including radar, Lidar, infrared sensors, inertial measurement units (IMUs), cameras, global positioning systems (GPS), etc., enhances the capability of ISAC networks for facilitating high-level environmental awareness, facilitating autonomous driving applications. However, multi-modal sensor fusion entails a complex process of transmitting and handling multi-source data, which remains a challenging open issue hinging on information sharing in ISAC networks.  

\subsubsection{Vehicular ISAC Networks}
In vehicular ISAC networks, collaboration facilitates the acquisition of information beyond the field of view of individual vehicles, thereby providing more comprehensive situational awareness for transportation purposes. Furthermore, the sensing results obtained by vehicular onboard sensors can be used for traffic coordination and platooning control. Moreover, given the growing population of vehicular radars, the development of a dedicated MAC tailored for radar sensing is a promising solution to improve sensing efficiency in ISAC networks, such as coordinating sensing tasks and routing radar information.

\subsection{Challenges} 

\subsubsection{Network Synchronization Requirements}
ISAC network synchronization poses a critical challenge in multi-user communication and multi-static sensing scenarios. In general, achieving accurate synchronization among ISAC transceivers at a clock level through high-precision timing protocols requires excessive signalling overhead. This is particularly crucial for uplink network sensing, mainly between mobile users and ISAC BSs. However, due to having limited communication resources, this can impose significant phase noise on both timing and carrier frequency offset compensation \cite{Liu2022Integrated}. Consequently, estimating target delay and Doppler frequency is contaminated.

\subsubsection{Limited Backhaul Constraints}
To implement cooperative transmission and sensing, all ISAC BSs in the cooperative cluster are linked to a central unit via backhaul connections to share the necessary information for cooperation. In many existing systems, the backhaul links are capacity-limited \cite{Ghimire2015Revisiting}, which is becoming a bottleneck for realizing the potential performance gain of both S\&C. To circumvent the limited backhaul capacity, a potential solution is to send pre-processed sensing results instead of sharing original signals, as detailed in Section \ref{DataFucion}. 

\subsubsection{Security and Privacy in Networked ISAC}
Having increased interactions between various ISAC transceivers within networks  inevitably results in security concerns. In particular, malicious passive sensing users could potentially exploit the waveforms transmitted from ISAC BSs to extract sensitive information about targets or their surroundings. Therefore, securing the sensing functionality within ISAC networks is crucial. In contrast to secure communication, ensuring security in sensing is more challenging due to the uncontrollability of the echo signals \cite{Bazzi2024Secure}. In general, single-cell ISAC offers limited DoF to improve performance while adhering to security constraints, particularly in situations where there is a high correlation between the channels of legitimate and illegitimate users. Joint BS selection and beamforming vector design in ISAC networks become a promising solution for significantly mitigating the data/sensing information leakage, a topic that has received limited attention in the literature. 

\begin{table*}[]
	\centering
	\small
	\vspace{-3mm}
	\caption{Different ISAC cooperation networks versus requirements.} 
	\label{Table2}
	\begin{tabular}{|ll|l|l|l|l|l|}
		\hline
		\multicolumn{2}{|l|}{\diagbox[]{\bf{\!\!\!Cooperation Level}}{\bf{Requirement\!\!\!}} }                     & {\bf{Information Sharing}} & {\makecell[c]{\bf{Transmission} \\{\bf{Latency}}}} &  {\makecell[c]{{\bf{Time}}\\{\bf{Synchronization}}}} & {\makecell[c]{\bf{Signaling} \\{\bf{overhead}}}} &  {\makecell[c]{\bf{Performance} \\{\bf{gain}}}} \\ \hline
		\multicolumn{2}{|l|}{Coordinated Cell Association}  &   {Target/user/BS state}   & Task level & No & Low & Low \\ \hline
		\multicolumn{2}{|l|}{Collaborative data fusion}                     & Estimated parameters  & Frame level  & No & Medium & Medium \\ \hline
		\multicolumn{2}{|l|}{Interference management}                     &  User/target CSI & No & No & Medium & High \\ \hline
		\multicolumn{1}{|l|}{\multirow{2}{*}{Cooperative S\&C}} & Non-coherent  & Data/echo signal &  \multirow{2}{*}{Frame level} & Symbol level & Very high & High \\ \cline{2-3} \cline{5-7} 
		\multicolumn{1}{|l|}{}   & Coherent  &  CSI and data/echo signal   &   & Phase level & Very high  & Very high \\ \hline
	\end{tabular}
\end{table*}
\begin{figure*}[t]
	\centering
	\includegraphics[width=14.5cm]{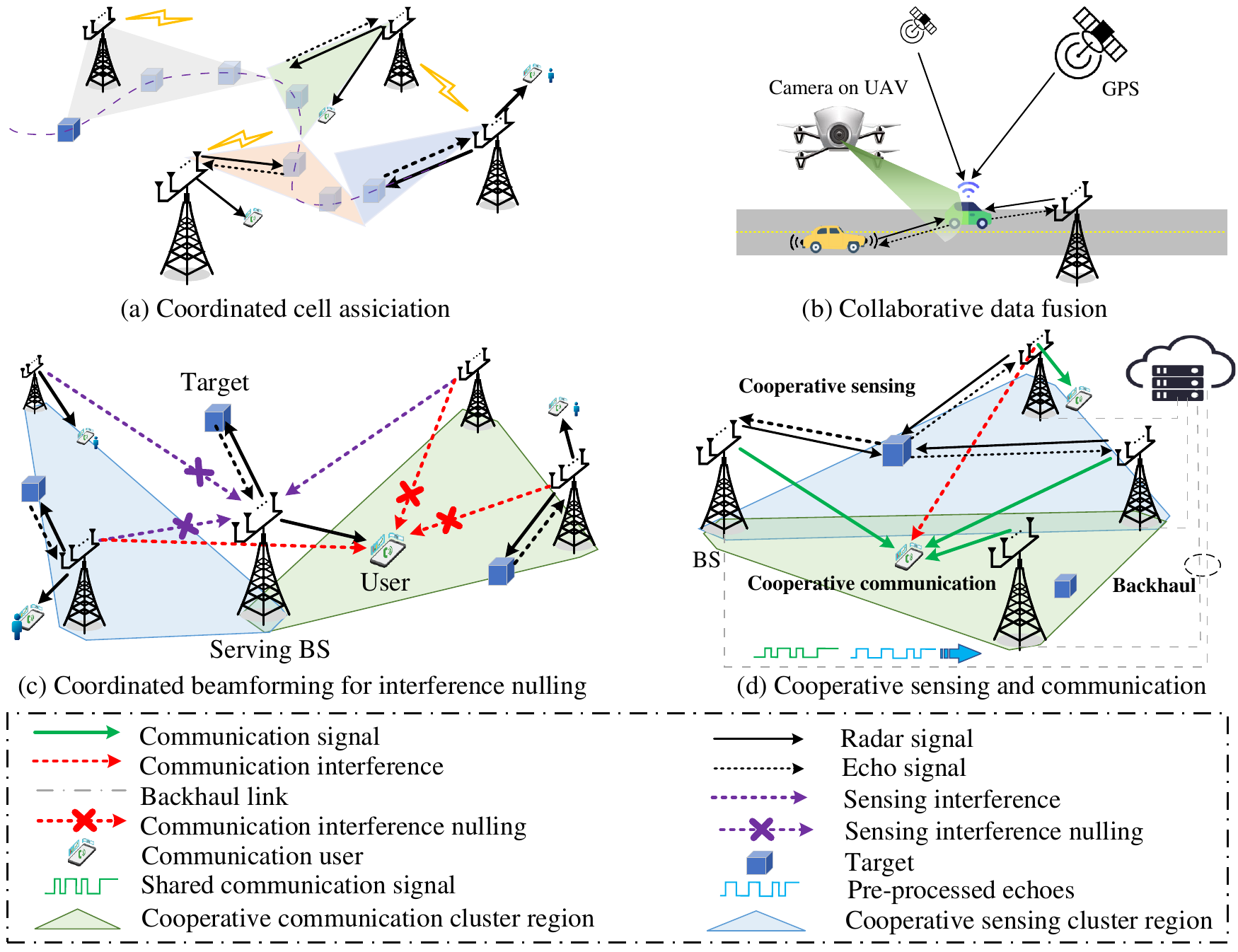}
	\vspace{-3mm}
	\caption{Cooperative S\&C networks.}
	\label{figure4}
\end{figure*}

\section{Cooperation Topology and Levels}
\label{DiffCooperations}
In this section, we categorize the collaboration of ISAC networks into four groups, and discuss their benefits, requirements, and challenges, as summarized in Table \ref{Table2}.

\subsection{Cooperation Clusters of BSs} 

In general, due to limited data transmission capacity among BSs, it is not feasible to engage all BSs in cooperative actions across the network. Consequently, determining which BSs partake in collaborative S\&C becomes pivotal. In contrast to BS clustering for communication-only networks, the cooperation cluster design of ISAC networks requires consideration of the networked S\&C cooperation framework, data fusion requirements, and performance balance between S\&C, as discussed  in Section \ref{Newconsideration}. Methods of partitioning BSs into cooperative groups typically encompass static clustering and dynamic clustering, both of which must take into account the integration gain of S\&C. Static clustering is less complex and imposes lower signalling overhead, but this method is not responsive to changes in the user and target locations, hence the performance gains remain limited. 
Dynamic clustering methods have been developed for accommodating changes in network and user/target mobility, such as new sites, sleeping cells, and load changes \cite{Meng2024Cooperative}. This scheme exacerbates both the scheduling and beamforming design complexity, which is a price to be paid for improving the performance by dynamically controlling the clusters for the sake of reducing the inter-cluster interference, thereby improving both the communication rate, the sensing accuracy, and energy efficiency. 

The size of the cooperative clusters is a crucial parameter for optimizing S\&C performance in ISAC networks when dividing BSs into multiple clusters. Small clusters may not fully benefit from the joint transmission and distributed radar, while larger clusters may result in excessive CSI feedback overhead and echo signal sharing \cite{meng2023network, Meng2024Cooperative}. Increasing the cluster size may improve the communication rate and sensing accuracy, albeit at the cost of additional signal processing and signalling. Furthermore, a larger cluster size may result in less energy efficiency.

\subsection{Cooperation Levels}
In this section, we present four distinct levels of cooperative schemes: coordinated cell association, collaborative data fusion, cooperative interference management, and joint cooperative S\&C arrangements.

\subsubsection{Coordinated Cell Association}
\label{CoordinationCellAssociaiton}
In general, multiple ISAC BSs should allocate and schedule S\&C tasks according to the geographical topology of ISAC networks and channel correlation between users and targets, thereby reducing interference and handover delay, improving resource utilization efficiency. Specifically, if a BS experiences high communication load and limited connectivity with other BSs, as exemplified by a BS in hot spots around shopping malls, its sensing traffic is inevitably reduced, since a high fraction of wireless resources is allocated to data transmission. Thus, the sensing node selection should also take into account the communication requirements. Moreover, from the perspective of each BS, the target and user having small angular separations or higher channel correlation coefficient should be served together at the same time for maximizing the overall spectral efficiency and coverage probability \cite{Liu2023Seventy}. On the other hand, it is preferable to track angularly separated targets falling into distinctiveness of each BS, thereby improving the sensing accuracy. In addition, when allocating S\&C tasks, the role of the BS in the network should be carefully considered. For instance, by allowing some BSs to perform dedicated sensing receiver tasks, network ISAC is capable of mitigating the self-interference in single-cell ISAC. 

Scenarios involving high-mobility users inside vehicles and targets across various BSs necessitate carefully crafted collaboration design to provide reliable S\&C services. However, how to handle the seamless handover of targets/users to ensure service reliability, while reducing signalling overhead is a challenging new issue \cite{Li2023TowardsSeamless}. For instance, in target tracking scenarios, if a target moves beyond the coverage of the currently assigned network entity or if the received echo signal is too weak for detection, collaborative involvement of another network entity becomes necessary for seamless S\&C services, as shown in Fig.~\ref{figure4}(a). Such handover decisions should be informed by factors such as the current or anticipated workload of each network entity and their potential coverage area in the direction of the moving target.

\subsubsection{Collaborative Data Fusion}
\label{DataFucion}
The sensing data gathered by individual radars may be overwhelmed by noise, fading, and interference. Collaborative data fusion combining measurements from multiple ISAC BSs mitigates uncertainties, thereby enhancing detection accuracy, reducing false positives, expanding coverage area, and improving the stability and reliability of perception in ISAC networks, as shown in Fig.~\ref{figure4}(b). Specifically, multiple ISAC BSs independently observe and estimate the parameters of the target, and fuse the target information by leveraging the coordinate system relationship among these BSs \cite{Lu2024Integrated}. For instance, in \cite{li2022multi}, a multi-point ISAC system was proposed that fuses the outputs from multiple ISAC devices for improving sensing performance by exploiting multi-view data redundancy, demonstrating beneficial fusion gain. During data sharing among BSs, communication beams can serve a dual purpose by concurrently facilitating additional sensing tasks, leading to improved data fusion efficiency. In addition, over-the-air computation can also be harnessed for significantly reducing the latency of sensing data fusion \cite{li2023over}. 

\subsubsection{Coordinated Beamforming for Interference Management}
\label{CoordinatedBeamforming}
The BSs share CSI information through dedicated backhaul links across the cooperative S\&C cluster. By adopting zero-forcing beamforming or minimum mean square error (MMSE) beamforming techniques, the beamforming policy can be designed for positioning the sensing beam within the null spaces of both the communication user channels and sensing receiver channels, as shown in Fig.~\ref{figure4}(c). This interference nulling scheme exploits extra spatial dimensions at the BS for creating spatial nulls at specific selective out-of-cell ISAC BS and/or user locations, thereby eliminating dominant interference for improving the S\&C performance. This can be accomplished without exchanging user and target information between the BSs, whilst relying solely on CSI. However, this approach requires additional BS antennas. In \cite{meng2023network}, it has been verified that interference-nulling substantially enhances both the average data rate and sensing accuracy. It is demonstrated that maximizing the ASE given by the product of the number of users served and the average throughput tends to allocate all spatial resources towards multiplexing and diversity gain, eschewing interference nulling. Conversely, in pursuit of sensing objectives, resource allocation leans towards suppressing direct BS-to-BS interference, for minimizing the impact on radar echo reception, particularly in case of numerous antennas, as inter-cell interference becomes a more dominant factor in affecting sensing performance.

\subsubsection{Cooperative Sensing and Communication}
\label{CooperativeSandC}
Both cooperative S\&C operations involve two types of signal-level cooperation schemes to convert hostile interference into useful signal. Explicitly, in communication, coherent transmission involves synchronized joint transmit precoding and coherent receiver combining. The practical challenges, such as the need for precise CSI feedback and stringent BS synchronization required for coherent cooperation, may constrain the potential gains. By contrast, non-coherent joint transmission does not require tight synchronization, where data is individually precoded from each cell. Distributed multiple-input multiple-output (MIMO) radar sensing can also be further classified into coherent and non-coherent distributed MIMO radar. Coherent processing leverages both the in-phase and quadrature-phase components of the transmitted signal, whereas non-coherent processing relies solely on the signal envelope. Strategically incorporating S\&C cooperation techniques shows considerable promise in attaining an improved and dynamically balanced performance within ISAC networks. For example, through the simultaneous utilization of CoMP joint transmission and distributed MIMO radar techniques, the authors of \cite{Meng2024Cooperative} proposed an innovative networked ISAC scheme, where multiple transceivers are employed for collaboratively enhancing both S\&C services, as shown in Fig.~\ref{figure4}(d). To strike a flexible tradeoff between the S\&C performance at the network level, \cite{Meng2024Cooperative} aims for optimizing the cooperative cluster sizes of S\&C under realistic backhaul capacity constraints. The expression of the CRLB in \cite{Meng2024Cooperative} derived for the localization accuracy reveals that harnessing $N$ ISAC transceivers enhances the average cooperative sensing performance across the network, in line with a scaling law of $\ln^2(N)$.
Crucially, this scaling law is less pronounced than the squared geometric gain of conventional distributed MIMO radar \cite{sadeghi2021target}, primarily due to the substantial pathloss from the distant BSs. This leads to reduced sensing performance gain.

\section{Synergies Between Network Sensing and Communication}
\label{SynergyBetweenSC}

\subsection{Network-Level Sensing-Assisted Communication}

In ISAC networks, BSs and sensors cooperatively explore radio wave transmissions, reflections, and scattering. This process facilitates the extraction of vital target- and environment-related information, such as traffic conditions and gathering of crowds, thereby enriching our awareness of the surrounding physical environment. As a result, utilizing information extracted from network sensing operations as prior knowledge may significantly enhance communication performance by predictive resource allocation and interference mitigation. For instance, by precisely localizing users and identifying surrounding blockages, network entities can optimize their transmission strategies for accommodating environmental variations, thereby reducing the overhead of beamforming design and improving spectrum efficiency \cite{Lu2024Integrated}. Additionally, by leveraging dynamic sensing results and urban environmental maps, resource allocation and deployment can be optimized for reliable S\&C performance based on predicting blockages. For example, by predicting a user's future movement trajectory and obstacle information, potential occlusion locations can be predicted, allowing for proactive handovers or resource adjustments to be made in advance.

\subsection{Network-Level Communication-Assisted Sensing}
The computing capabilities of numerous sensing devices are limited by their hardware and cost, which may impair their ability to promptly extract sensing information. Due to the strict timeliness requirements of data processing, it can be challenging for BSs or mobile sensors having limited computational capabilities to perform sensing tasks in practice, particularly in delay-sensitive ISAC missions, such as target tracking. In these scenarios, network entities having ISAC capabilities can offload demanding intensive sensing tasks to powerful edge servers. This offloading process accelerates data processing, enabling low-latency extraction of sensing information. In this case, the topological information of ISAC networks can improve sensing efficiency, particularly when the network is not fully connected. 

In addition, addressing the tight latency requirements of sensing information within the constraints of communication rates poses significant practical challenges, especially for wireless data aggregation in dense BS scenarios. A promising strategy for reducing data fusion delay is constituted by over-the-air computation techniques \cite{li2023over}. The core concept of over-the-air computation is to leverage the waveform superposition property of a wireless channel, enabling simultaneous aggregation of data transmitted by multiple BSs or sensors. Based on such a mechanism, the signals simultaneously transmitted by BSs or sensors are superposed over the air and aggregated at the server through weighted summation, with the weights representing channel conditions \cite{li2023over}. Consequently, latency can be significantly reduced by supporting simultaneous S\&C.

\subsection{Mutual Benefits of Networked Sensing and Communication}
When the aforementioned reciprocity principles between network S\&C mechanisms are simultaneously applied, S\&C tasks can be specifically designed for assisting each other in achieving win-win integration. The results of network sensing can assist in the allocation of communication resources. On the other hand, the performance of communication network may impact the fusion of sensory data. This raises an intriguing question regarding the optimal resource allocation of ISAC networks when maximizing S\&C performance, since enhancing any of the individual sensing or communication performances can indeed improve the bottom-line performance. For example, prioritizing resources towards communication in a sensing-assisted communication scenario will indeed improve the bottom-line communication performance. However, dedicating more resources to sensing naturally improves the sensing operation, which in turn may also benefit the bottom-line communication performance. To address this issue, a plausible approach is comparing the gradient of performance gain attained by mutual assistance. Specifically, if the performance improvement achieved by increasing the sensing resources exceeds the degradation caused by the reduction of communication resources, more sensing resources should be allocated to the ISAC network. Otherwise, the communication performance gain attained by sensing cannot compensate for the performance erosion caused by the time and power consumption. Under this scenario, opting for allocating more communication resources is the most favourable choice. 

\section{Case Study: Interference Management and Cooperation Scheme}
\label{CaseStudy}
 
\begin{figure}[t]
	\centering
	\includegraphics[width=7.5cm]{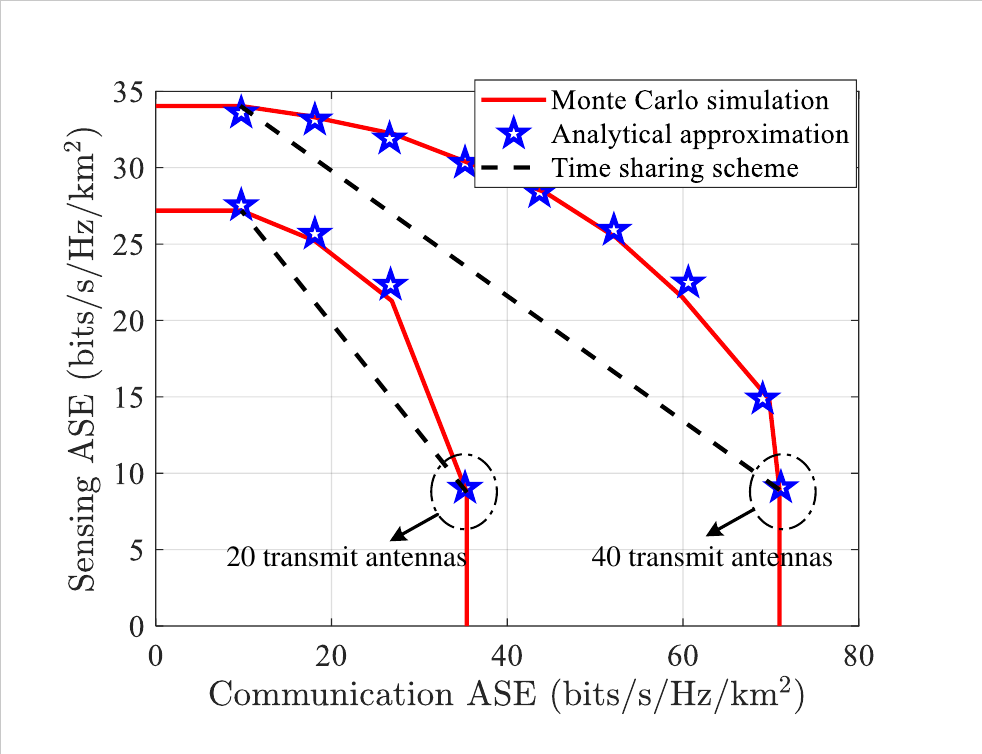}
	\vspace{-3.5mm}
	\caption{Area spectral efficiency tradeoff between S\&C versus different service quality constraints.}
	\label{figure5}
\end{figure}

To demonstrate the efficiency of cooperative schemes in ISAC networks, we investigate two cases: interference nulling and cooperative S\&C by Monte Carlo simulations for a BS density of $1/{\rm{km}}^2$ and user/target density of $20/{\rm{km}}^2$, respectively. The system parameters are as follows: The number of receive antennas is $5$, the transmit power is $1$W at each BS, the average RCS $1$, the noise power is $- 80$dB. The carrier frequency is 5 GHz, and the bandwidth is 50 MHz. Initially, we harness coordinated beamforming in cooperative S\&C BS clusters for interference nulling, as discussed in Section \ref{CoordinatedBeamforming}.
In this approach, each BS transmits independent data to several users, while concurrently carrying out sensing tasks for several targets, such as localization and recognition, utilizing unified ISAC signals. To individually mitigate S\&C interference, we select $Q$ nearest BSs for interference nulling purposes in sensing, and $L$ nearest BSs for that in communication \cite{meng2023network}, as shown in Fig.~\ref{figure4}(c). The improvement of S\&C performance hinges on optimizing the number of users and targets served and the size of cooperative BS clusters. This joint optimization enhances the ASE, defined as the product of the number of users (or targets) and the average spectral efficiency for communication (or sensing). To demonstrate the effectiveness of the cooperative ISAC scheme under various setups, we compare it to a time sharing scheme established by connecting two corner sensing-communication performance points, as shown in Fig.~\ref{figure5}. The sensing and communication ASE regions for the cooperative ISAC scheme exhibit a notable expansion compared to the time sharing schemes as the number of transmit antennas increases. This expansion is attributed to improved Degrees of Freedom (DoF) and optimal spatial resource allocation, enhancing multiplexing gain, diversity gain, and interference nulling capabilities.

\begin{figure}[t]
	\centering
	\includegraphics[width=7.5cm]{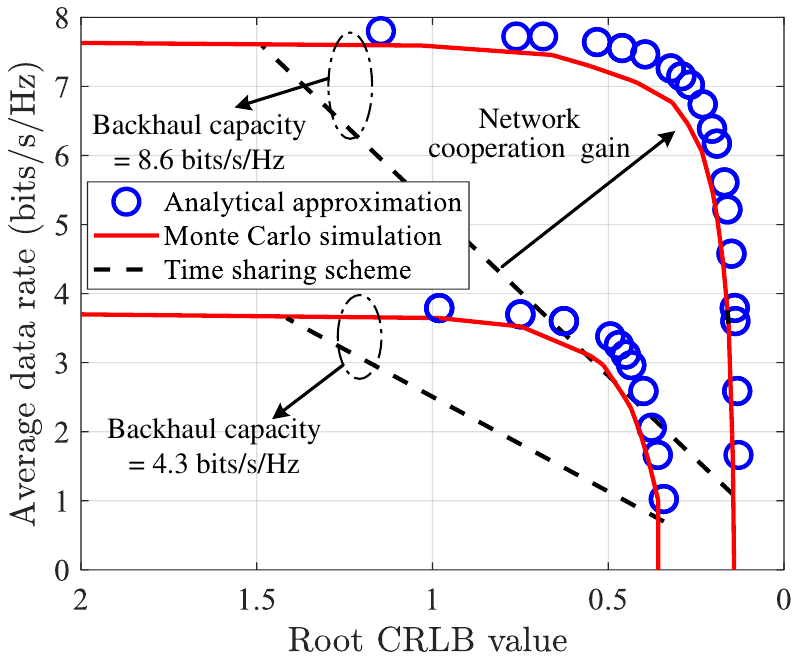}
	\vspace{-3mm}
	\caption{Performance tradeoff between average data rate and root CRLB versus different backhaul capacity constraints.}
	\label{figure6}
\end{figure}

Subsequently, we introduce a cooperative S\&C scheme for ISAC networks with $4$ transmit antennas of each BS, where a cluster of cooperative BSs collaboratively transmit the same communication data to the served user, while multiple BSs form distributed multistatic MIMO radars for collaboratively sensing each target, as shown in Fig.~\ref{figure4}(d). To unlock the full potential of cooperative S\&C within network-level ISAC frameworks, as discussed in Section \ref{CooperativeSandC}, it's crucial to jointly optimize the transmit power of S\&C signals and the size of cooperative BS clusters \cite{Meng2024Cooperative}. As illustrated in Fig. \ref{figure6}, as the backhaul capacity increases, the performance boundaries of S\&C expand significantly. This expansion is attributed to the availability of more feasible cooperation resources attained by larger cooperative S\&C cluster sizes. A fundamental tradeoff emerges between the average data rate and the average CRLB of the entire network. Moreover, as depicted in Fig. \ref{figure6}, compared to the time sharing scheme, the attainable performance region of the optimal cooperative scheme significantly expands upon increasing the backhaul capacity. This is attributed to the increased capacity of backhaul links, facilitating effective coordination of transmit power and multi-cell resources within the network, leading to higher gains in cooperative cluster design for S\&C. 

\section{Future Extensions}
\label{Extensions}
\subsection{Smart Propagation Engineering for ISAC Networks}
To further improve the ISAC service quality, ISAC networks can work in conjunction with smart propagation engineering technologies, e.g., intelligent surfaces, fluid antenna systems, unmanned aerial vehicles (UAVs), etc. Specifically, intelligent surfaces enhance S\&C signal coverage and signal quality by appropriately adjusting the phase shifts to reduce interference, while fluid antenna systems create new DoFs to strike a balance between S\&C performance by optimizing antenna positions. This will unveil the full capability of radio signals by evolving to smart radio control by exploiting a suite of new features, such as self-aggregation and self-configuration. Furthermore, these smart propagation engineering techniques can provide higher mutual assistance gain between S\&C. 

\subsection{Semantically aware ISAC networks}
Describing the transmit signal and target parameters in a semantic form may hold the promise of reducing the latency and backhaul capacity requirements, thereby facilitating multi-cell ISAC cooperation. However, the exchange of semantic information between different ISAC BSs requires new protocols and functions, such as semantic extraction, semantic composition, and semantic instruction. These functions must seamlessly interact with the radio layer, often through operator functions responsible for controlling network nodes. Providing a unified framework of semantically aware ISAC networks is non-trivial due to the various S\&C task categories and complex interactions in cooperative ISAC networks.

\subsection{Self-adaptive AI in ISAC Networks}
ISAC networks are capable of real-time active sensing and automatic updating of Artificial Intelligence (AI) networks, potentially resulting in self-adaptive AI, where AI systems can adjust their behaviour and learning process based on the evolving wireless environments without human intervention. Continuous real-time sensory data collection during communications provides valuable opportunities for adaptive resource management and predictive network configurations. in the face of uncertainty, creating native network intelligence. A valuable research avenue is to explore self-adaptive AI to facilitate multi-BS collaboration within the ISAC network, thereby improving training efficiency and facilitating mutual assistance between S\&C.

\section{Conclusions}
\label{Conclusions}
Network-level ISAC has the potential of improving S\&C performance by leveraging the network-level DoF. We discussed new design metrics, and opportunities, and highlighted the essential challenges in ISAC networks. We then revealed the mutual benefits of networked S\&C and presented several cooperation levels along with the corresponding requirements. Finally, we validated the effectiveness of cooperative ISAC schemes through simulations. 


\footnotesize  	
\bibliography{mybibfile}
\bibliographystyle{IEEEtran}

\vspace{3mm}
\noindent {\bf{Kaitao Meng}} is a Marie Curie Fellow with the Department of Electronic and Electrical Engineering, University College London, U.K. He is a Guest Editor for the IEEE TCCN. He served as the Co-Chair of the IEEE GlobeCom 2024 Workshop and Track Chair of the IEEE VTC-spring 2025.

\vspace{3mm}
\noindent {\bf{Christos Masouros}} (FIEEE, FAAIA) is a full professor at University College London, U.K. He received the 2023 IEEE ComSoc Stephen O. Rice Prize, the 2021 IEEE SPS Young Author Best Paper Award, Best Paper Awards at IEEE GlobeCom 2015 and IEEE WCNC 2019. He is an Editor for IEEE TWC, IEEE OJ-SP, and IEEE OJ-COMS. He has been an Editor for IEEE TCOM, IEEE CL, and a Guest Editor for a number of IEEE JSTSP and IEEE JSAC issues. 

\vspace{3mm}
\noindent {\bf{Athina Petropulu}} is Distinguished Professor of Electrical and Computer Engineering at Rutgers University. She is Fellow of IEEE and the American Association for the Advancement of Science. She was 2022-2023 President of the IEEE Signal Processing Society and 2009-2011 Editor-in-Chief of the IEEE TSP. For further information please see {\url{https://en.wikipedia.org/wiki/Athina_Petropulu}}.

\vspace{3mm}
\noindent {\bf{Lajos Hanzo}} (FIEEE'04) received Honorary Doctorates  from the Technical University of Budapest (2009) and Edinburgh University (2015). He is a Foreign Member of the Hungarian Science-Academy, Fellow of the Royal Academy of Engineering (FREng), of the IET, of EURASIP and holds the IEEE Eric Sumner Technical Field Award. For further details please see {\url{http://www-mobile.ecs.soton.ac.uk}}, {\url{https://en.wikipedia.org/wiki/Lajos_Hanzo}}.

	\end{document}